\documentclass[usenatbib]{mn2e}
\pdfoutput=1
\usepackage{graphicx}
\usepackage{mathptmx}
\usepackage{pdfsync}
\usepackage{amsmath}
\usepackage{amssymb}
\usepackage{aas_macros}
\usepackage[usenames,dvipsnames]{color}
\usepackage[a4paper,total={17.8cm,24.0cm},centering]{geometry}
\usepackage[pdfpagelabels]{hyperref}
\hypersetup{pdfauthor={Michael J. Williams et al.},pdftitle={The
stellar kinematics and populations of boxy bulges: cylindrical rotation
and vertical gradients},breaklinks=true}

\newcommand{\refsec}[1]{Section~\ref{#1}}
\newcommand{\reffig}[1]{Fig.~\ref{#1}}

\newcommand{\kms}{km\,s$^{-1}$}

\newcommand{\reff}{\hbox{\ensuremath{R_\mathrm{e}}}}

\newcommand{\oiii}{[O\,\textsc{iii}]}

\newcommand{\ZH}{[\ensuremath{\mathrm{Z}/\mathrm{H}}]}
\newcommand{\aFe}{[\ensuremath{\alpha/\mathrm{Fe}}]}

\title[Kinematics and populations of boxy bulges]{The stellar kinematics
and populations of boxy bulges: cylindrical rotation and vertical
gradients\thanks{Based on observations collected at the European
Organization for Astronomical Research in the Southern Hemisphere, Chile
(programmes 64.N-0545, 65.N-0126 and 66.B-0073)}}

\author[Michael J. Williams et al.]{Michael J.
Williams$^{1,2}$\thanks{Email: williams@astro.ox.ac.uk}, Michel A.
Zamojski$^3$, Martin Bureau$^1$, Harald Kuntschner$^2$,
\newauthor Michael R. Merrifield$^4$, P. Tim de Zeeuw$^{2,5}$, 
Konrad Kuijken$^5$
\\$^1$Sub-Department of Astrophysics, University of Oxford, Denys
Wilkinson Building, Keble Road, Oxford OX1 3RH, UK
\\$^2$European Southern Observatory, Karl-Schwarzschild-Str. 2, D-85748
Garching bei M\"unchen, Germany
\\$^3$Spitzer Science Center, California
Institute of Technology, Pasadena, CA 91125, USA
\\$^4$ School of Physics and Astronomy, University of Nottingham,
Nottingham NG7 2RD, UK
\\$^5$ Sterrewacht Leiden, Universiteit Leiden, Postbus 9513, 2300 RA
Leiden, the Netherlands}

\begin{document}

\date{Accepted 2011 February 11. Received 2011 February 11; in original
form 2010 October 20}

\pagerange{\pageref{firstpage}--\pageref{lastpage}} \pubyear{2010}

\maketitle
\label{firstpage}

\begin{abstract} 
Boxy and peanut-shaped bulges are seen in about half of edge-on disc
galaxies. Comparisons of the photometry and major-axis gas
and stellar kinematics of these bulges to simulations of bar formation
and evolution indicate that they are bars viewed in projection. If the
properties of boxy bulges can be entirely explained by assuming they are
bars, then this may imply that their hosts are pure disc galaxies with no
classical bulge. A handful of these bulges, including that of the Milky
Way, have been observed to rotate cylindrically, i.e. with a mean
stellar velocity independent of height above the disc. In order to
assess whether such behaviour is ubiquitous in boxy bulges, and whether
a pure disc interpretation is consistent with their stellar populations,
we have analysed the stellar kinematics and populations of the boxy or
peanut-shaped bulges in a sample of five edge-on galaxies. We placed
slits along the major axis of each galaxy and at three offset but
parallel positions to build up spatial coverage. The boxy bulge of
NGC~3390 rotates perfectly cylindrically within the spatial extent and
uncertainties of the data. This is consistent with the metallicity and
$\alpha$-element enhancement of the bulge, which are the same as in
the disk. This galaxy is thus a pure disc galaxy. The boxy bulge of
ESO~311-G012 also rotates very close to cylindrically. The boxy bulge of
NGC~1381 is neither clearly cylindrically nor non-cylindrically rotating,
but it has a negative vertical metallicity gradient and is
$\alpha$-enhanced with respect to its disc, suggesting a composite bulge
comprised of a classical bulge and bar (and possibly a discy
pseudobulge). The rotation of the peanut-shaped bulge of NGC~5746 is
difficult to classify, but the peanut-shaped bulge of IC~4767 does not
rotate cylindrically. Thus, even this relatively small sample is
sufficient to demonstrate that boxy bulges display a range of rotational
and population properties, indicating that they do not form a
homogeneous class of object.
\end{abstract}

\begin{keywords}
galaxies:~abundances --- galaxies:~kinematics~and~dynamics ---
galaxies:~bulges --- galaxies:~stellar content
\end{keywords}

\section{Introduction}

The central regions of disc galaxies host bulges whose nature (or
absence) must be explained by any model of galaxy formation and
evolution. A bulge is a physical protrusion of stars above and below the
plane of the disc and/or an excess of light above the inwards
extrapolation of the radial exponential profile of the disc. There are
three principal observational classes of bulges: classical bulges,
pseudobulges, and boxy or peanut-shaped bulges
\citep[see e.g.][]{Athanassoula:2005}.

Classical bulges have dynamical and photometric properties and stellar
populations that are similar to elliptical galaxies of the same mass
\citep{Kormendy:1982,Davies:1983,Emsellem:2004,Kormendy:2004,Falcon-Barroso:2006,Thomas:2006,MacArthur:2009}
and are thought to be the end products of the same formation processes:
either monolithic collapse (e.g. \citealt*{Eggen:1962};
\citealt{Larson:1974,Carlberg:1984,Pipino:2004}) or multiple
hierarchical mergers \citep[e.g.][]{White:1978,Cole:1994,Thomas:1999}.

\begin{figure*}
\includegraphics[width=8.4cm]{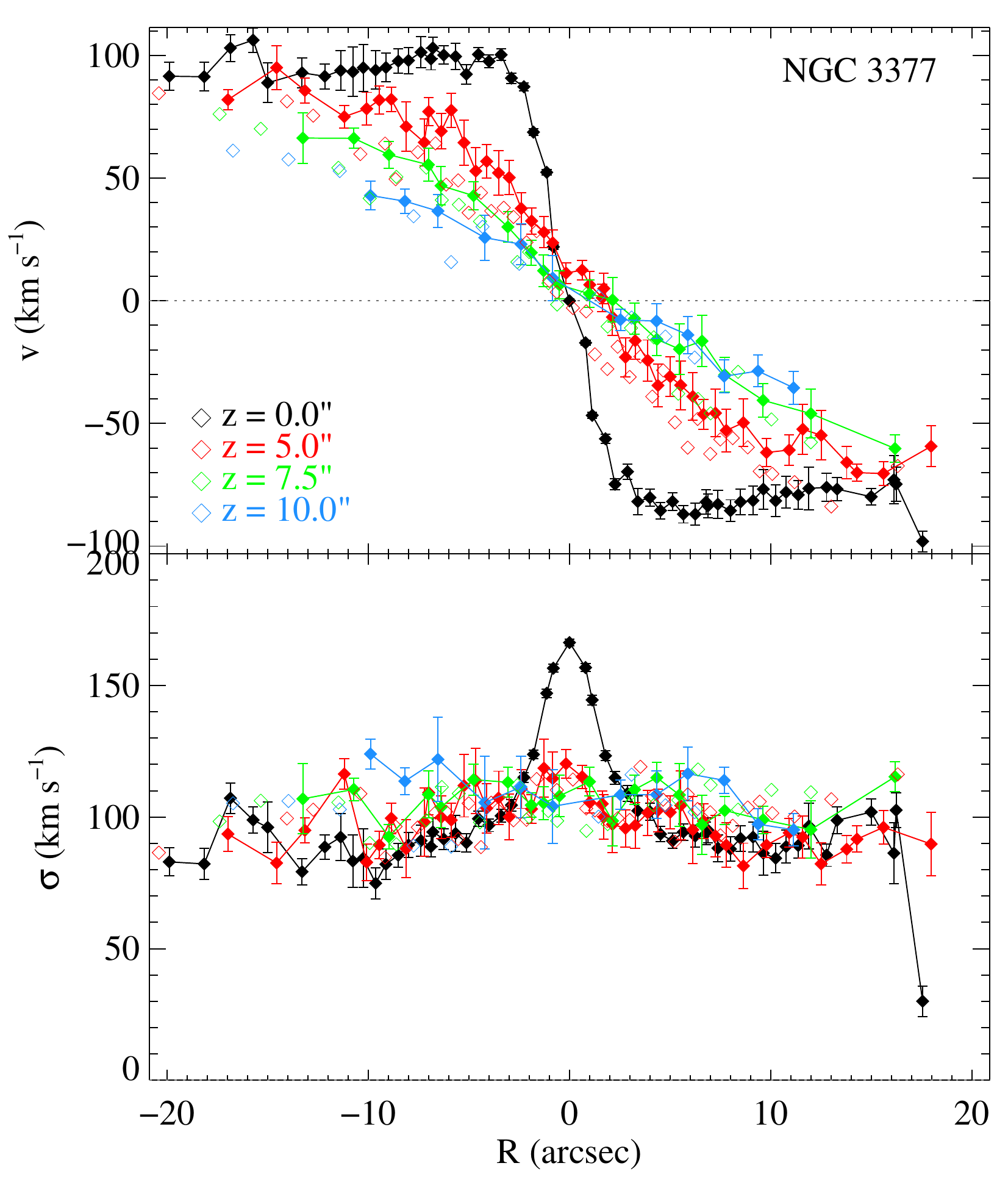}
\includegraphics[width=8.4cm]{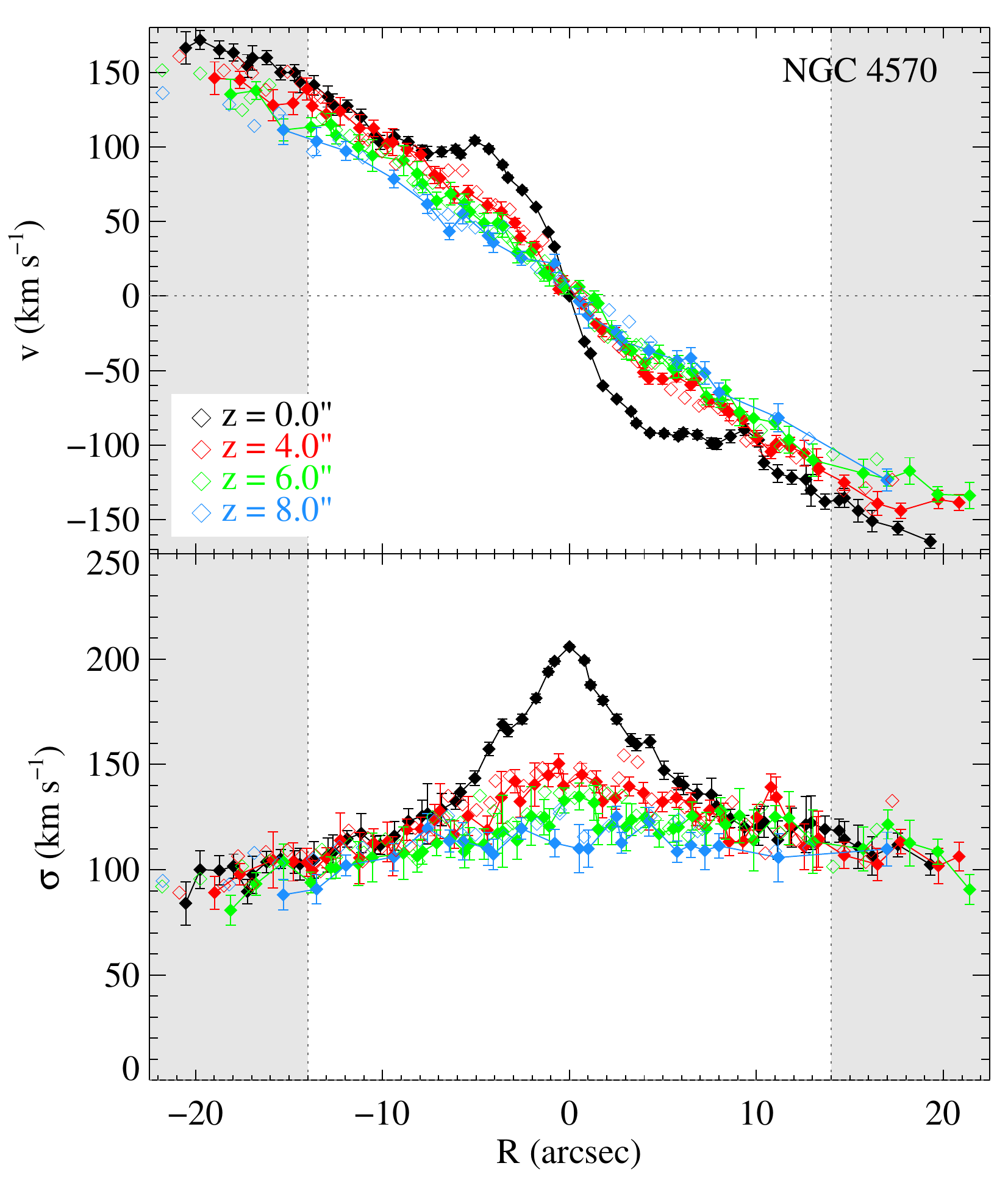}
\caption{Stellar line-of-sight rotation curves and velocity dispersion
profiles for two non-cylindrically rotating edge-on galaxies from the
SAURON sample (the fast rotator elliptical NGC~3377 and the classical
bulge of the S0 NGC~4570), constructed by extracting pseudoslits from
the integral-field kinematics presented in \protect\cite{Emsellem:2004}.
The stellar kinematics along the major axis are shown in black, and
those from pseudoslits offset from, but parallel to, the major axis are
shown in order of increasing distance from the major axis in red, green
and blue. The offset distance $z$ is shown. Data from one side of the
major axis are shown as filled diamonds (joined by lines), and data from
the other side as open diamonds (with error bars omitted for clarity).
The vertical dotted lines at $R = 14\arcsec$ for NGC~4570 indicate the
effective radius \reff, i.e. the approximate extent of the bulge (beyond
which the figure is shaded grey). The effective radius of NGC~3377 is
38\arcsec, i.e. beyond the radial range shown. Note that, within each
bulge at a given $R$, $v$ falls systematically with height above the
major axis, i.e. the rotation pattern is not cylindrical.}
\label{fig:sauron}
\end{figure*}

Pseudobulges have embedded spiral structure, disc-like near-exponential
light profiles, flattened axial ratios, and rotation-dominated dynamics
\citep[see][and references therein]{Kormendy:2004}. They are thought to
be secularly rearranged disc material that was driven inwards by bars,
ovals and possibly spiral arms. The observational constraints on the
stellar populations of pseudobulges are less strong because studies have
focussed on bigger and brighter classical bulges, decomposing their
extended star formation history is more complicated than in classical
bulges and ellipticals, and dust is a significant problem for the
late-type discs in which pseudobulges are more usually found.
Nevertheless, the available population data are not strongly
inconsistent with the secular model
(\citealt{Peletier:1999,Gadotti:2001}, but see the pilot sample of
\citealt{MacArthur:2009}).

An important third class of bulge, and the focus of this paper, is the
boxy and peanut-shaped bulges, which are present in about half of
edge-on disc galaxies \citep{Lutticke:2000}. Their major-axis stellar
and gas kinematics
\citep{Kuijken:1995,Merrifield:1999,Bureau:1999,Chung:2004,Mendez-Abreu:2008}
and light profiles \citep{Bureau:2006} are consistent with simulations
of the formation and buckling of galactic bars
\citep{Combes:1990,Raha:1991,Athanassoula:1999,Bureau:2005} and point to
them being bars viewed in projection. Peanut-shaped bulges are thought
to be bars whose orientation is exactly perpendicular to the
line-of-sight, while boxy bulges are seen in galaxies whose bar is at an
intermediate angle. Bars oriented exactly parallel to the line-of-sight
do not appear boxy or peanut-shaped. In that sense, boxy and
peanut-shaped bulges are not the axisymmetric features normally thought
of as `bulges'. Since the dominant model is that they are bars, which
are not the end products of collapse or merger, but rather redistributed
disc material, boxy bulges are also sometimes referred to as
pseudobulges. However, for the purposes of clarity in the present paper,
we avoid this, since boxy bulges are truly thick.

In the absence of transformational mergers, bars, which are found in
two-thirds of disc galaxies
\citep{Eskridge:2000,Whyte:2002,Marinova:2007}, are expected to play a
crucial and perhaps dominant role in disc galaxy evolution
\citep[e.g.][]{Sellwood:1993,Kormendy:2004,Hopkins:2010}. Boxy and
peanut-shaped bulges therefore provide a unique perspective on a feature
that is crucial for understanding galaxy evolution. By observing them
above the plane of the disc, one can determine the intrinsic properties
of bars in a way that is almost free of thin disc light pollution.
Moreover, the observed dynamics of bars as a function of height above
the plane is a crucial additional constraint on models of the nature of
bulges, including ruling out the presence of a dynamically hot classical
bulge/merger remnant. For example, \cite{Shen:2010} modelled the
line-of-sight mean velocity and velocity dispersion of M giant stars
observed by the BRAVA survey in the Milky Way bulge \citep{Howard:2009}.
They placed an extremely low upper limit on the mass of any hot spheroid
component, $\lesssim$7\%; i.e. the Milky Way appears to be a pure disc
galaxy (but see \citealt{Babusiaux:2010}). Giant pure disc galaxies
present an acute challenge for cosmological simulations of galaxy
formation
\citep[e.g.][]{van-den-Bosch:2001,Carollo:2007,Dutton:2009,Kormendy:2010a}.

Perhaps the most unusual property of dynamical models of bars viewed in
projection is that, within the bulge, the rotational velocity does not
change with height \citep[e.g.][]{Combes:1990,Athanassoula:2002}. That
is to say, lines of constant velocity are parallel to each other and
perpendicular to the major axis, such that $\partial\,v/\partial\,|z|
\sim 0$, where $v$ is the line-of-sight velocity and $z$ is along the
minor axis of the (edge-on) galaxy. This unusual behaviour was first
observed in the boxy bulge of NGC~4565 by \cite{Kormendy:1982}, who
named it `cylindrical rotation'. It has since been noted in the
following eight galaxies, all of which have a bulges that is, more or
less, boxy: NGC~128 \citep{Jarvis:1990}, NGC~3069 \citep{Shaw:1993},
NGC~1055 \citep{Shaw:1993a}, NGC~4442 \citep{Bettoni:1994}, NGC~7332
\citep{Fisher:1994,Falcon-Barroso:2004}, NGC~4220 and NGC~4425
\citep{Falcon-Barroso:2006}, and the Milky Way \citep{Howard:2009}. In
simulations, all bulges that rotate cylindrically are boxy or
peanut-shaped, but not all boxy or peanut-shaped rotate cylindrically
\citep{Athanassoula:2002}. Cylindrical rotation is not present (or has
not been reported) in simulated unbarred galaxies, or in observations of
classical bulges, where rotational velocity falls systematically with
height within the bulge (see \reffig{fig:sauron} for two illustrative
examples from the SAURON sample; \citealt{de-Zeeuw:2002,Emsellem:2004}). 

In our search of the literature, we found no observations of the absence
of cylindrical rotation in a boxy bulge. One of the goals of the present
targeted study of boxy bulges is to determine whether this is
confirmation bias, or whether cylindrical rotation is truly ubiquitous
in boxy bulges and therefore a requisite property of all realistic
models of bars. The extent or absence of cylindrical rotation in boxy
bulges would be of interest because it might constrain the concentration
of the dark haloes \citep{Athanassoula:2002}.

Our second goal is to examine the stellar populations of boxy bulges.
The null hypothesis might be that, since bars are secularly rearranged
disc material (rather than material accreted or formed during rapid
mergers), the metallicities of boxy bulges should be similar to those of
their host discs. If, however, the metallicities of boxy bulges differ
from those of their discs, or if they have metallicity gradients, this
is not necessarily a problem for the bar buckling scenario. As pointed
out by \cite{Freeman:2008}, the stars that have been scattered furthest
from the disc are the oldest stars, and therefore formed from the least
metal-enriched fuel. The buckling process may therefore establish a
negative minor-axis metallicity gradient. Indeed, such behaviour is
observed in the Milky Way \citep{Zoccali:2008} and in NGC~4565, the
archetypal boxy, cylindrically-rotating bulge \citep{Proctor:2000}. The
strength of the metallicity gradient established in this way must be a
function of both the rate of enrichment in the disc and the timescale
over which the bar buckles and heats the disc. The process must also
establish a corresponding positive age gradient. We know of no
chemodynamical simulations of this process and, in particular, no
prediction of the extent to which minor-axis gradients in boxy bulges
are expected to differ from those in classical bulges. 

Moreover, we know of no prediction regarding the relative abundance of
$\alpha$-elements in boxy bulges. The naive expectation, however, is
that the \aFe\ ratio of a boxy bulge should match that of its host disc.
It is difficult to understand how secularly rearranged disc material
could be segregated such that a boxy bulge of disc origin is
$\alpha$-enhanced with respect to its disc. Enhanced \aFe\ ratios
indicate that material was $\alpha$-enriched by core collapse (type II)
supernovae before type Ia supernovae Fe-enrichment, which implies a
formation timescale $\lesssim$ 1\,Gyr. Bulges that are $\alpha$-enhanced
with respect to their disc therefore contains stars
formed on much shorter timescales than the disc stars (and not just at
earlier times). If observed in a galaxy, this would suggest that its
bulge is at least partially comprised of material formed in a merger or
rapid collapse, or material of external origin. We discuss some of the
issues that complicate this naive expectation in \refsec{sec:discussion}

In this paper we present a focussed spectroscopic analysis of the boxy
bulges of five edge-on disc galaxies. In \refsec{sec:observations} we
describe the sample, the data and their reduction. We present the derived
stellar kinematics in \refsec{sec:kinematics} and the stellar populations in
\refsec{sec:populations}. We summarise our results and suggest future
work in \refsec{sec:discussion}.

\begin{figure}
\includegraphics[width=8.4cm]{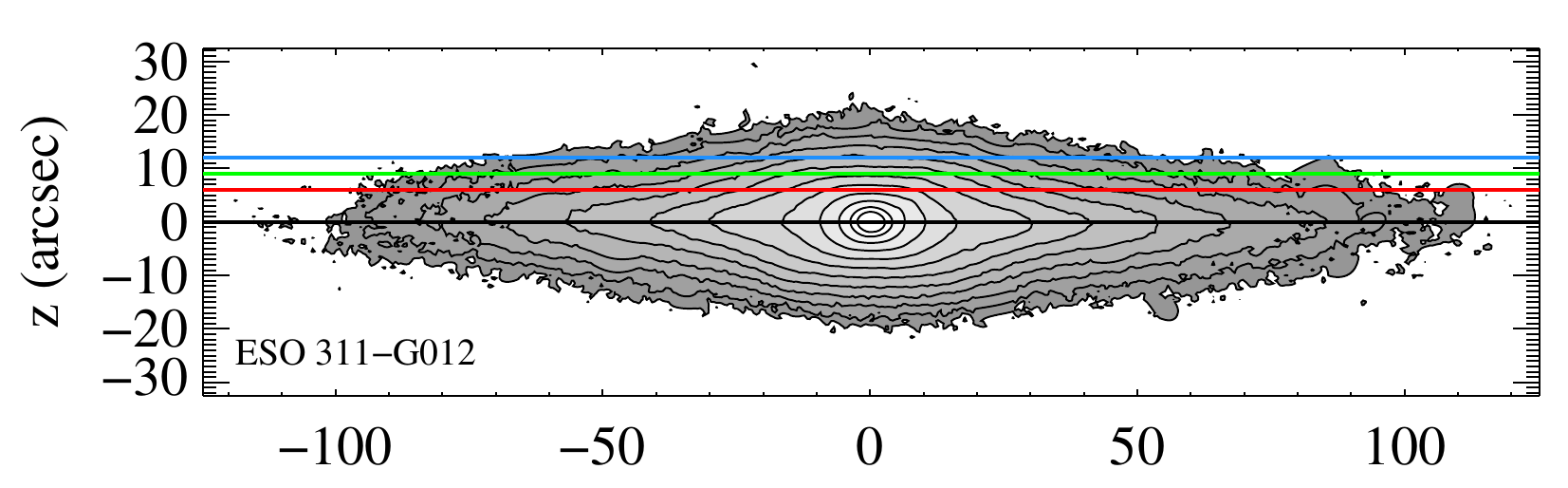}
\includegraphics[width=8.4cm]{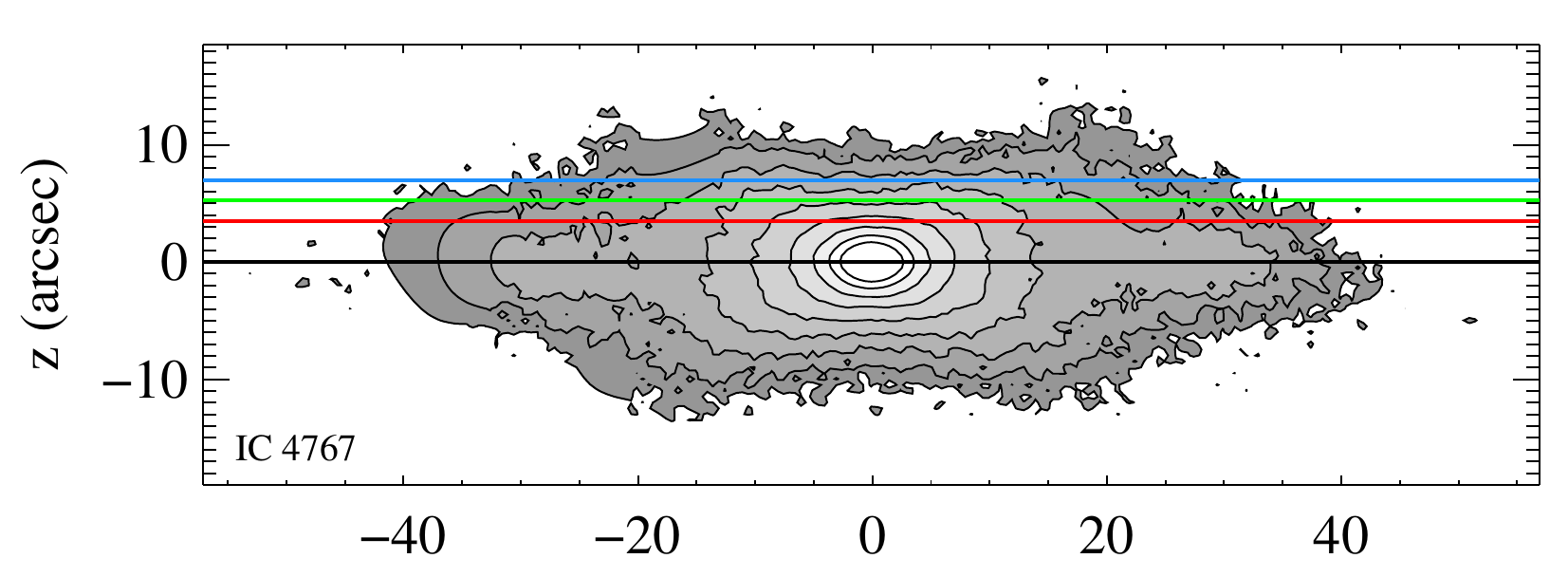}
\includegraphics[width=8.4cm]{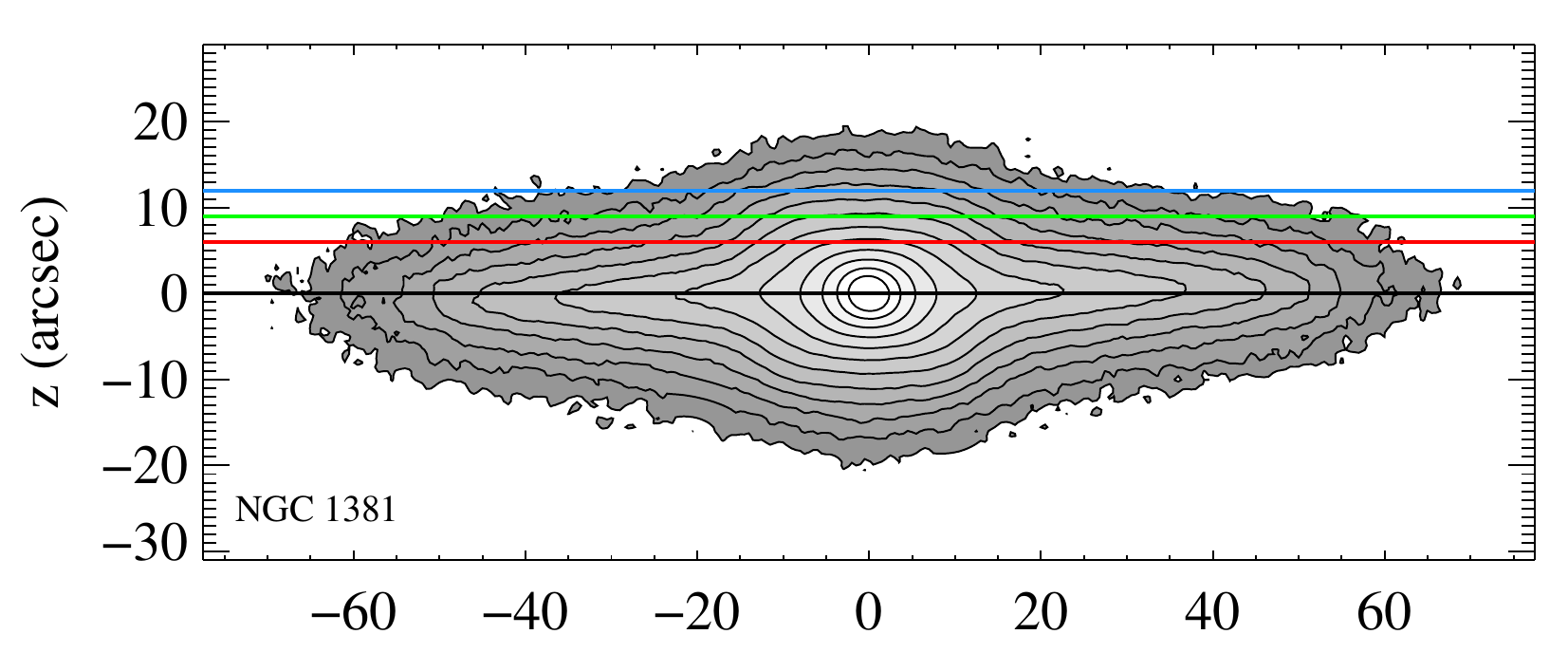}
\includegraphics[width=8.4cm]{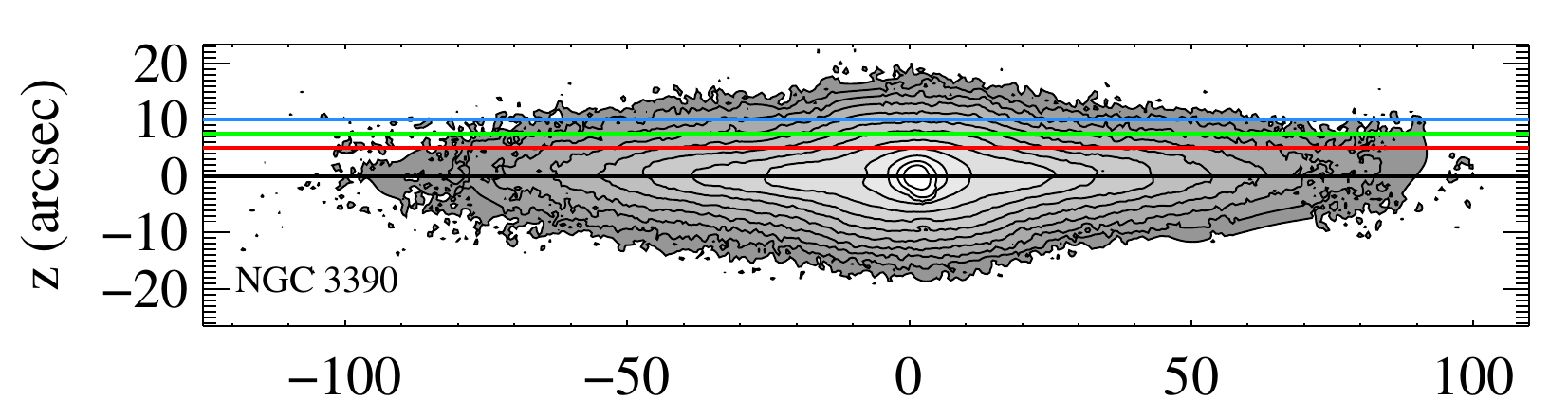}
\includegraphics[width=8.4cm]{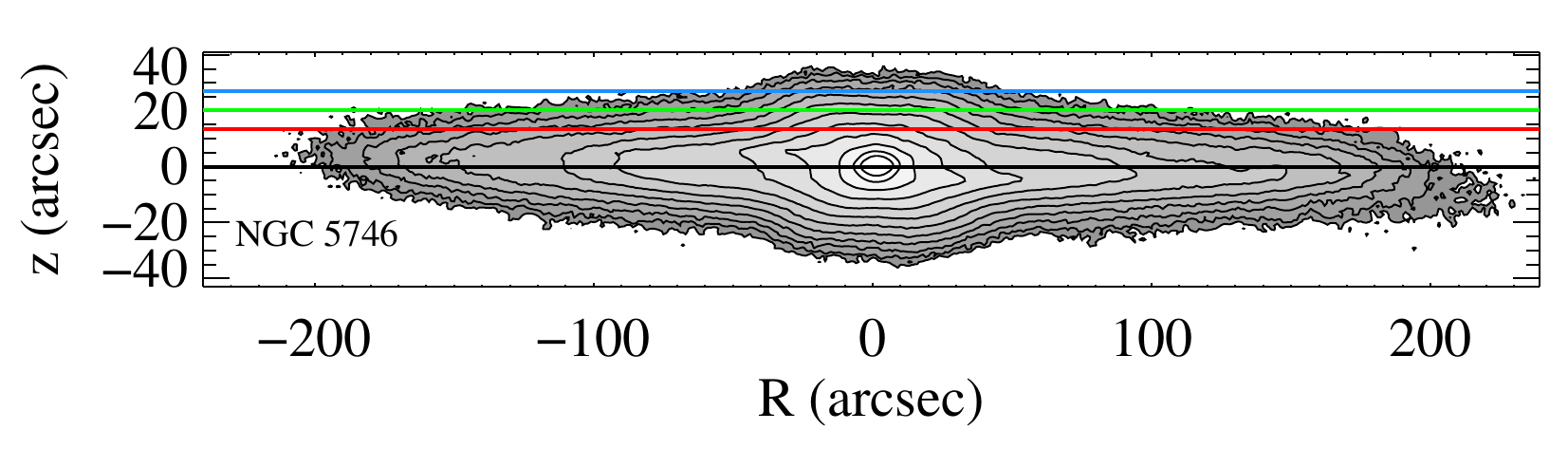}
\caption{$K$-band images of the sample galaxies with the slits overlaid.
Contours are separated by 0.5\,mag. Images taken from
\protect\cite{Bureau:2006}.}
\label{fig:images}
\end{figure}

\section{Observations and data reduction}
\label{sec:observations}

\begin{table*}
\caption{Sample properties and slit positions.}
\label{tab:sample}
\begin{tabular}{lllcclll}
\hline
Galaxy & RC3 & Bulge type & Bulge extent & \multicolumn{3}{c}{Offset slit positions} \\
       & classification & & (arcsec) & \multicolumn{3}{c}{(arcsec)} \\
\hline
ESO 311--G012  & S0/a?    & Boxy             & 15     &  6.0   &  9.0  & 12     \\
  IC 4767     & S0+\^     & Peanut           & 18     &  3.5   &  5.25 & 7      \\
  NGC 1381    & SA0       & Boxy             & 10     &  6.0   &  9.0  & 12     \\
  NGC 3390    & Sb        & Boxy             & 10     &  5.0   &  7.5  & 10     \\
  NGC 5746    & SAB(rs)b? & Peanut           & 20     & 13.5   & 20.25 & 27     \\
\hline
\end{tabular}
\begin{minipage}{\textwidth}
\emph{Notes.} Bulge extent is defined here as the radius of the maximum
scale height of the disc, as measured from the vertical photometric
parametrizations presented in Bureau et al. (in preparation). The offset
slit positions are the distance each slit was offset from the major axis
of the galaxy.
\end{minipage}
\end{table*}

Our sample consists of five galaxies whose bulges were classified as
boxy or peanut-shaped by \cite{Lutticke:2000} using Digitized Sky Survey
images, and whose major-axis stellar/gas kinematics and stellar
structure have been homogeneously studied
\citep{Bureau:1999,Chung:2004,Bureau:2006}. All of the galaxies are
edge-on or nearly edge-on. NGC~5746 is the furthest from edge-on in the
sample. It is inclined at 85$^\circ.$5, measured by fitting an ellipse to
the prominent ring visible in near-infrared \emph{Spitzer} 
Infrared Array Camera images (PI: Giovanni Fazio). 

We observed the sample using the New Technology Telescope's ESO
Multi-Mode Instrument (EMMI) in long-slit spectroscopy mode. The
observations were made in 2000 January, May and October (programmes
64.N-0545, 65.N-0126 and 66.B-0073). We positioned the slit along the
major axis and then parallel to it at three offset positions. In this
way we built up spatial coverage similar to that of a sparse, wide-field
integral-field spectrograph. We present the sample properties and slit
positions in Table~\ref{tab:sample}. We show the slits superimposed on
images of the sample in \reffig{fig:images}.

We observed three other galaxy bulges (NGC~1055, NGC~1247 and NGC~7123),
two of which are not boxy, during the runs  but we dropped them from the
present analysis because the data are poor or incomplete, and the SAURON
project and other work have superseded any contribution these
observations could have made to our understanding of the dynamics and
populations of classical bulges
\citep[e.g.][]{Proctor:2002,Emsellem:2004,Thomas:2006,Falcon-Barroso:2006}.

We reduced the data using standard techniques in IRAF (Image Reduction
and Analysis Facility), yielding flat-fielded, wavelength-calibrated
sky-subtracted two-dimensional spectra at four slit positions for each
galaxy. Exposure times ranged from 30 minutes for the major axis slits
to up to 3 hours for the slits furthest from the disc. The reduced
two-dimensional spectra were linearly binned in wavelength and cover the
range $4830$--$5470$\,\AA. The spectral resolution of the spectra is
1.0\,\AA\ full-width half-maximum (equivalent to $\sigma_\mathrm{inst}
\approx 25$\,\kms{} at 5150\,\AA{}). The spatial axis has a pixel scale
of 0.9\,arcsec/pixel. We determined the central bin (i.e. the spectrum
for which the cylindrical radius $R = 0$) using a S\'ersic fit to the
profile of each spectrum. This measurement is correct at the sub-arcsec
level for the major axis slit, but is uncertain by $\lesssim 5$\,arcsec
away from the midplane, where the radial profile is noisy and weakly
peaked. The central bin for each slit may therefore be a point that does
not lie exactly on the minor axis. This uncertainty may introduce a
small, systematic relative offset between the kinematic profiles derived
for each slit.

Spectrophotometric flux standards were not observed during the observing
runs, so that we are unable to precisely flux calibrate our data.
However, as described in \refsec{sec:populations}, this does not
significantly affect our ability to measure line strengths.

\section{Stellar kinematics}
\label{sec:kinematics}

\subsection{Methods}

\begin{figure*}
\includegraphics[width=8.4cm]{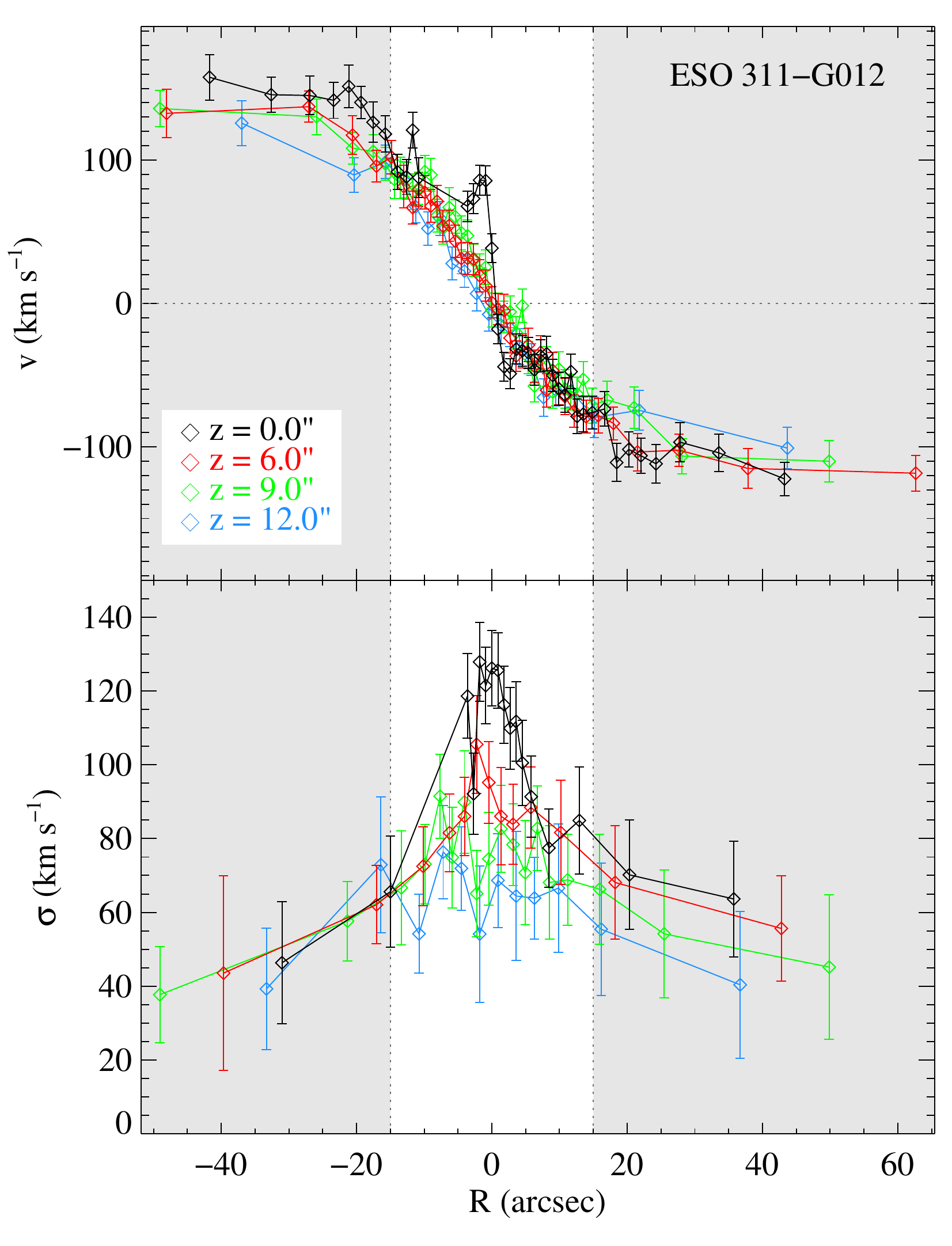}
\includegraphics[width=8.4cm]{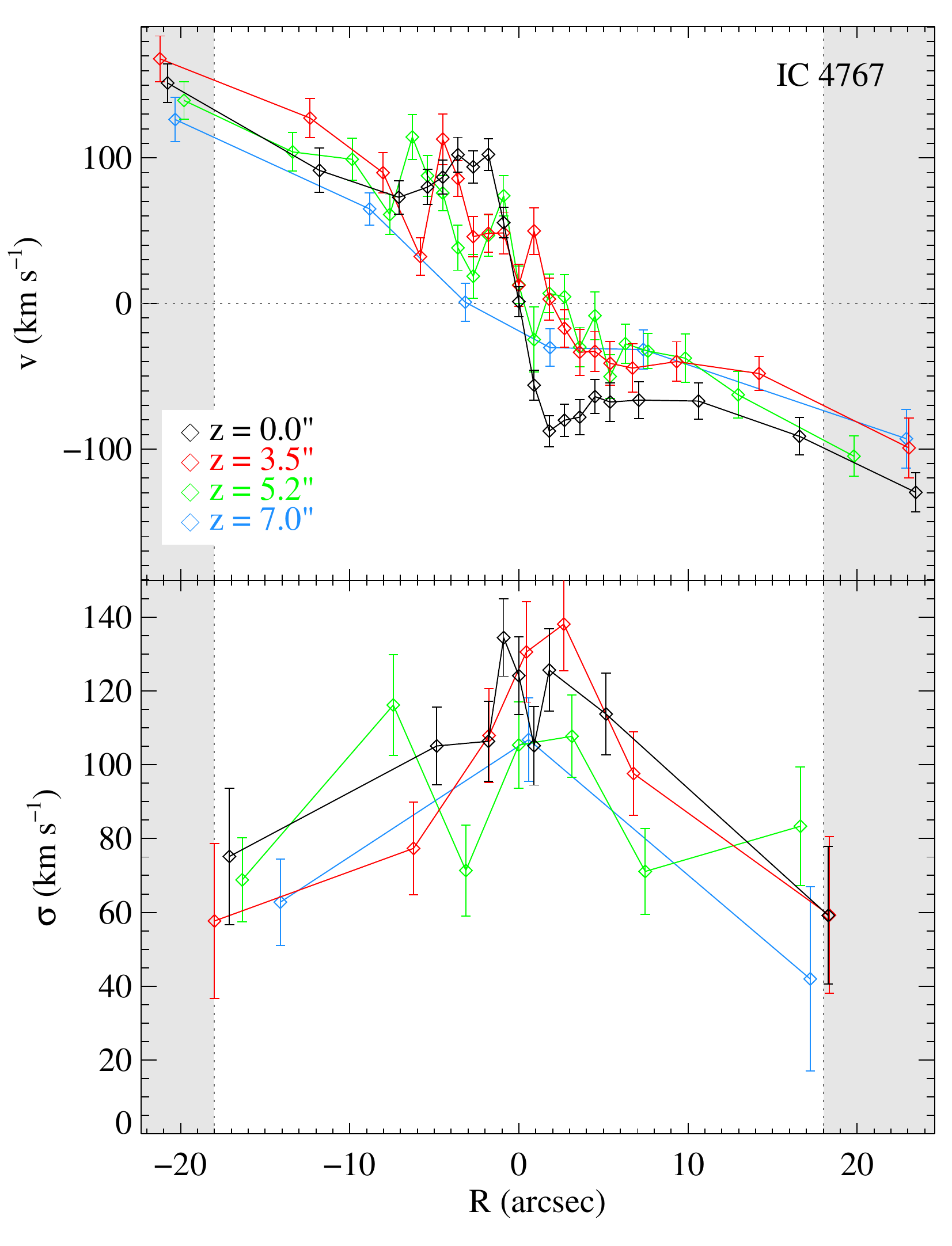}
\includegraphics[width=8.4cm]{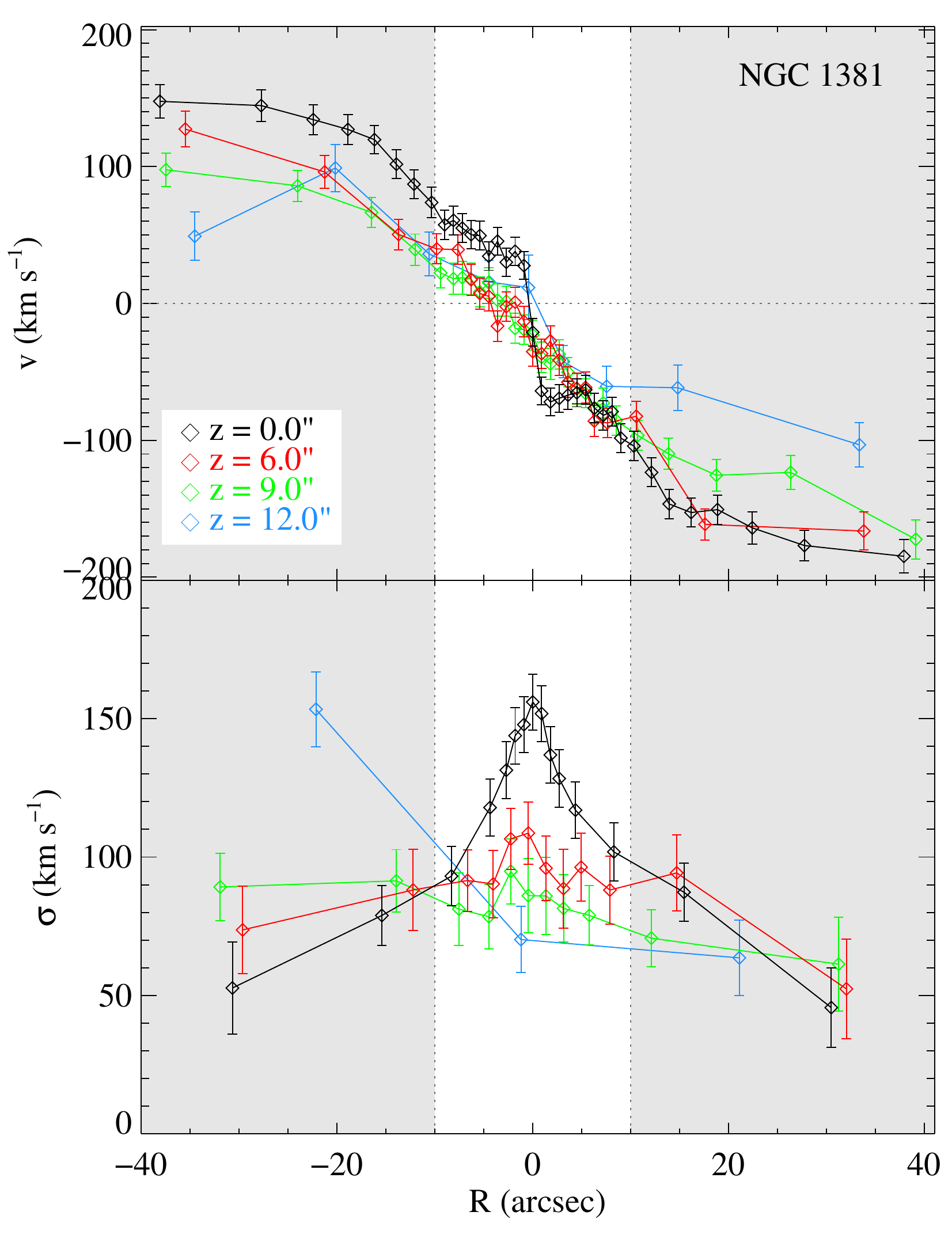}
\includegraphics[width=8.4cm]{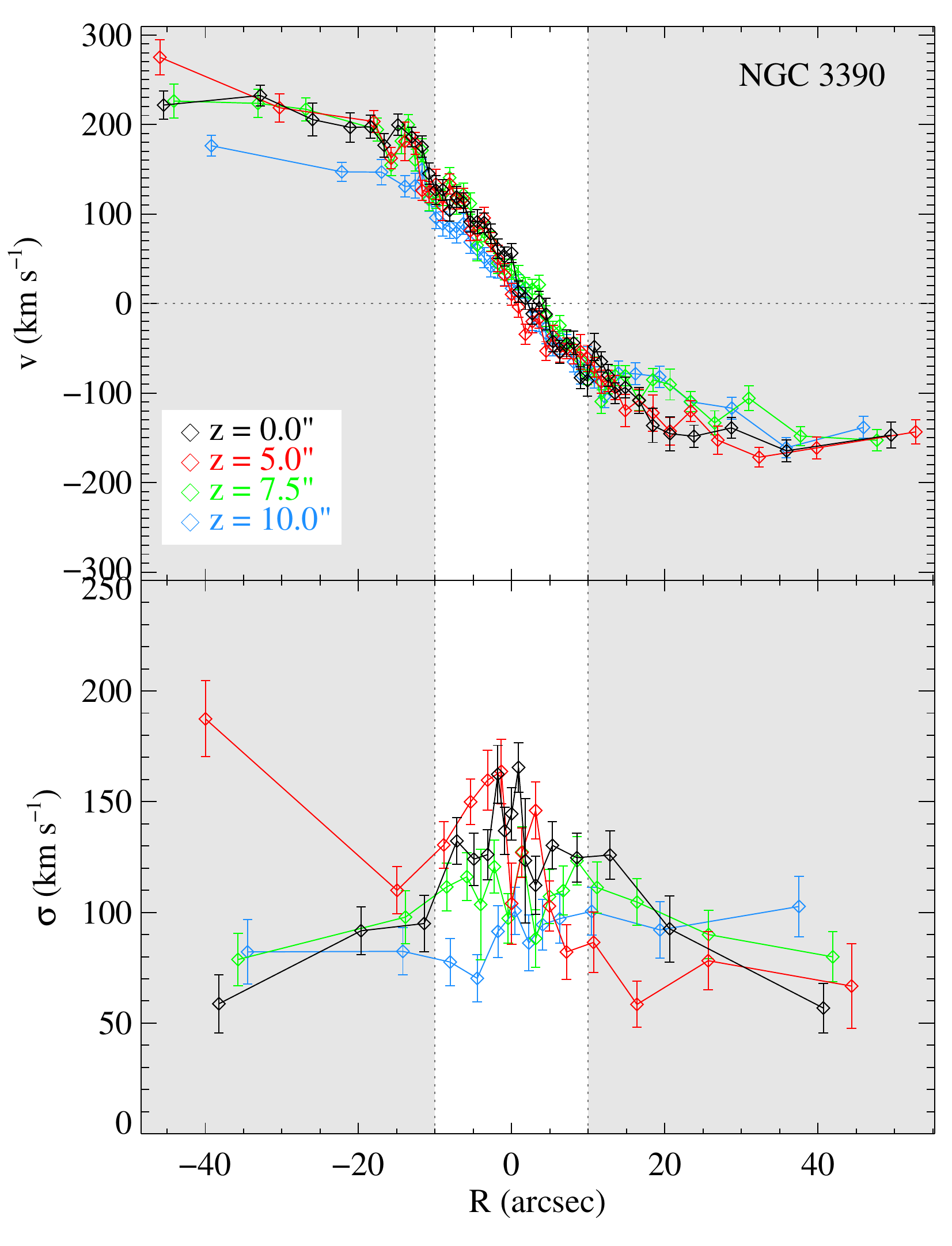}
\caption{Stellar line-of-sight mean velocity (top) and velocity
dispersion (bottom) for the five sample galaxies. The stellar kinematics
of the major axis are shown in black, and those of the offset slits are
shown in order of increasing distance from the major axis in red, green
and blue. See Table~\ref{tab:sample} and \reffig{fig:images} for the
slit positions. The vertical dotted lines indicate the approximate
extent of the bulge (see Table~\ref{tab:sample}). The shaded regions at
large radii therefore correspond roughly to the galaxy disc. The same
systemic velocity has been subtracted from all points, determined from a
pPXF fit to the total major axis spectrum.}
\label{fig:kinematics}.
\end{figure*}

In order to determine the stellar kinematics, we first binned the data
spatially to enhance the signal-to-noise ratio $(S/N)$. To measure the
line-of-sight mean velocity $v$, we typically binned to a $S/N$ of about
10 per \AA, but lowered this to 5 per \AA\ for a few of the most offset
slits. The outermost bin in each slit often does not reach the desired
$S/N$ level, but we analysed it (and included in these figures) for
completeness. To measure the velocity dispersion $\sigma$, we rebinned
the data to twice the $S/N$, sacrificing spatial resolution.

We extracted absorption line stellar kinematics from the binned spectra
using penalized pixel fitting (pPXF; \citealt{Cappellari:2004}). All
galaxies except NGC~1381 show clear signs of H$\beta$ or \oiii\ emission
in their major-axis spectrum, but the data in each bin are of
insufficient $S/N$ to constrain the emission line properties and
subtract the emission in a physically motivated way using, e.g., GANDALF
\citep{Sarzi:2006}. To derive the absorption line stellar kinematics,
the appropriate spectral regions are therefore simply
masked.\footnote{Note that we do treat emission properly when
determining the absorption line strengths in \refsec{sec:populations},
where we bin to much higher $S/N$.} We also masked a series of bad
columns on the CCD at a rest wavelength of $\approx$5050\,\AA.

We used a subset of 88 stars from the MILES library of 985 observed
stellar spectra as templates \citep{Sanchez-Blazquez:2006}. The MILES
spectra have a spectral resolution of 2.3\,\AA\, while our spectra have
a resolution of 1.0\,\AA\, so we first degraded our spectra to the MILES
resolution. In principle, the penalized pixel fitting algorithm can
robustly recover dispersions down to around half of this degraded
spectral resolution, i.e. $\approx$28\,\kms. Using the Fourier
Correlation Quotient method \citep{Bender:1990} and a template star with
a spectral resolution of $\approx$32\,\kms, \cite{Chung:2004} measured
dispersions $> 50$\,\kms{} along the major axis of all five galaxies in
the present sample. Since the true dispersions of these galaxies is of
order the spectral resolution of the MILES library, working at the MILES
spectral resolution should not affect our results for the line-of-sight
velocity dispersion. To verify this, we repeated the kinematic
extraction for NGC~1381 working at the spectral resolution of our
observations. To do this we used degraded template spectra from the
ELODIE library of 1388 stars (whose spectra have a resolution of
12\,\kms, i.e. higher than that of our observations;
\citealt{Prugniel:2001}). Within the uncertainties, the $v$ and $\sigma$
radial profiles we obtained agree with those recovered using the MILES
library, with no sign of any systematic disagreement.

Our data are of insufficient $S/N$ to constrain $h_3$ (skewness) and
$h_4$ (kurtosis) using a Gauss-Hermite parametrization of the
line-of-sight velocity distribution (LOSVD). We therefore enforce a pure
Gaussian LOSVD. Our conclusions regarding cylindrical rotation, which
depend only on $v$, are not affected by this decision.

We estimated the uncertainties on the derived kinematics by running pPXF
on 100 random realisations of each spectrum. The realisations were
constructed by adding wavelength-independent Gaussian noise assuming 
Poisson statistics. To these uncertainties we added uncertainties due to
wavelength calibration, which we estimate to be $\lesssim 0.2$\,\AA,
i.e. $\lesssim$10\,\kms. These wavelength calibration uncertainties
generally dominate the uncertainties in $v$, but are less important for
$\sigma$, where the spectral noise dominates.

\subsection{Results}

\begin{figure}
\includegraphics[width=8.4cm]{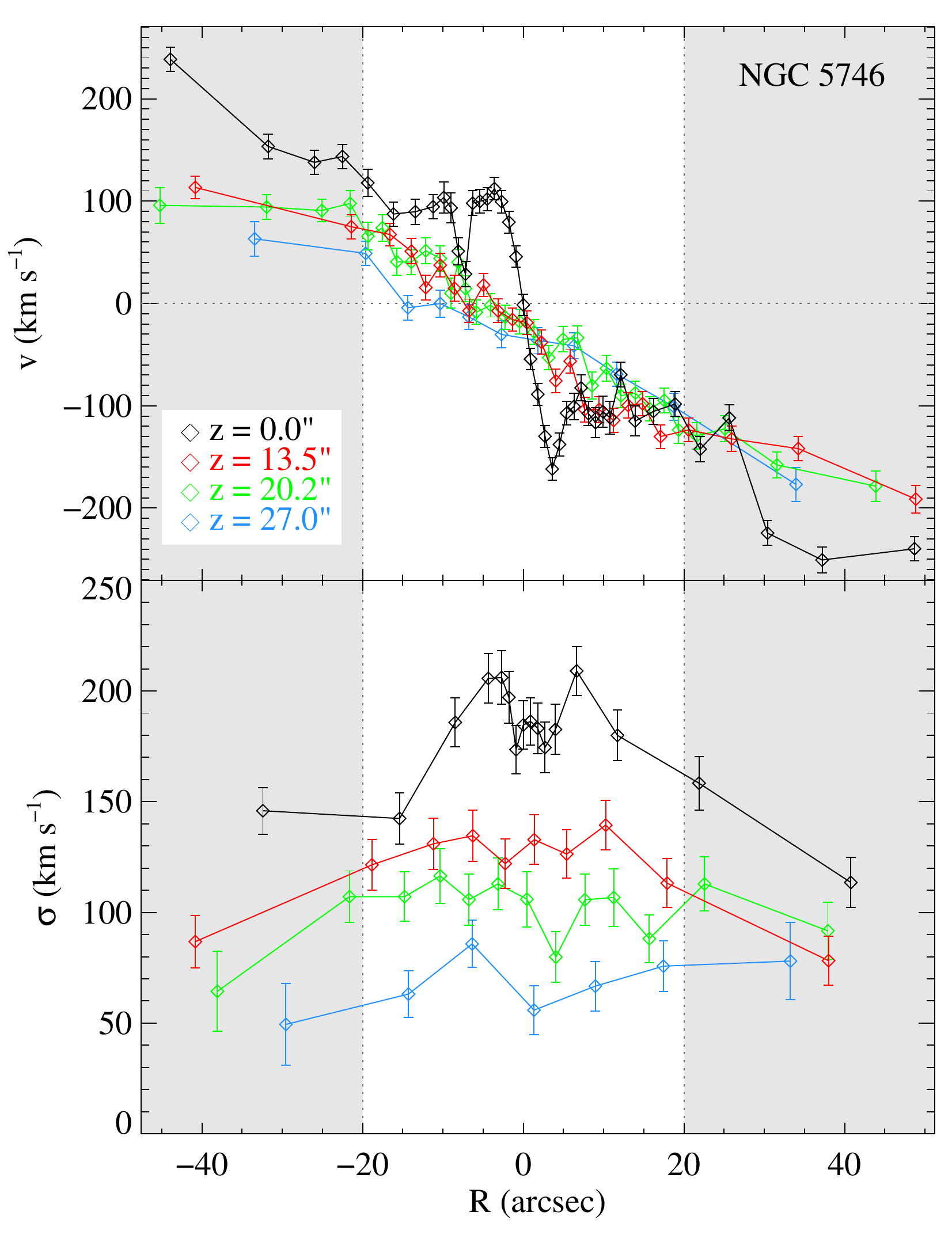}
\addtocounter{figure}{-1}
\caption{--- continued.}
\end{figure}

The line-of-sight mean stellar velocity $v$ and velocity dispersion
$\sigma$ calculated as described above are presented in
\reffig{fig:kinematics}. The kinematics of the major axis slit are shown
in black and those of the offset slits are shown in order of increasing
distance from the major axis in red, green and blue. In each plot, we
indicate the approximate extent of the bulge with dotted vertical lines.

The boxy bulge of NGC~3390 is a clear example of cylindrical
rotation. Within the uncertainties, it shows no systematic variation
of rotational velocity as a function of height at all. 

The behaviour of the boxy bulge of ESO~311--G012 is close to
cylindrical. The major-axis rotation is somewhat discrepant from that of
the offset slits; the rotation curve along the plane of the disc is
`double-humped', while those of the offset slits are not. This does not
mean that the boxy bulge is not rotating cylindrically. The strong
double-hump feature is seen in simulations of barred galaxies
\citep[e.g.][]{Bureau:2005} and is likely the dynamical signature of bar-driven
inward transport of material \citep[e.g.][]{Kormendy:2004}, which leads
to the formation of a flattened round central excess of material, i.e. a
pseudobulge. The distinguishing feature of cylindrical rotation is
therefore that $v$ does not vary systematically with height away from
the major axis, which is true of ESO~311--G012. Taken together, we
interpret the presence of a double hump in the rotation curve along the
major axis and cylindrical rotation in the offset slits as consistent
with simulations of the formation and dynamics of bars
\citep[][]{Combes:1990,Athanassoula:2002,Bureau:2005}. 

The classification of the type of rotation in the boxy bulge of NGC~1381
is marginal. The velocity along the major axis is extremely discrepant
from that along the offset slits. The slits at 6\arcsec\ and 9\arcsec\
are corotating, but the slit at 12\arcsec\ is of extremely low $S/N$,
and is binned to a correspondingly coarse spatial resolution. While this
galaxy lacks the clear systematic decrease in $v$ with increasing $z$
seen in the examples in \reffig{fig:sauron}, it is not a clean case of
cylindrical rotation either.

The same is true of the peanut-shaped bulge of NGC~5746, but the picture
is further complicated here by the rather irregular rotation curves
within the bulge. This may be due to the prominent dust `lane' in
NGC~5746, which, because this galaxy is a few degrees away from edge-on
obscures a fraction of the galaxy. Strong absorption may cause the local
mean line-of-sight velocity to differ significantly from the azimuthal
velocity at the tangent point along the line-of-sight, and the extent of
this difference may vary in a complicated way with distance along the
minor axis.

The peanut-shaped bulge of IC~4767 is not cylindrically rotating. At
negative $R$, even accounting for the uncertainties and the irregularity
of the rotation curve, $v$ falls systematically with $z$. At positive
$R$, the signature of non-cylindrical rotation is less clear.

\begin{figure*}
\includegraphics[width=\textwidth]{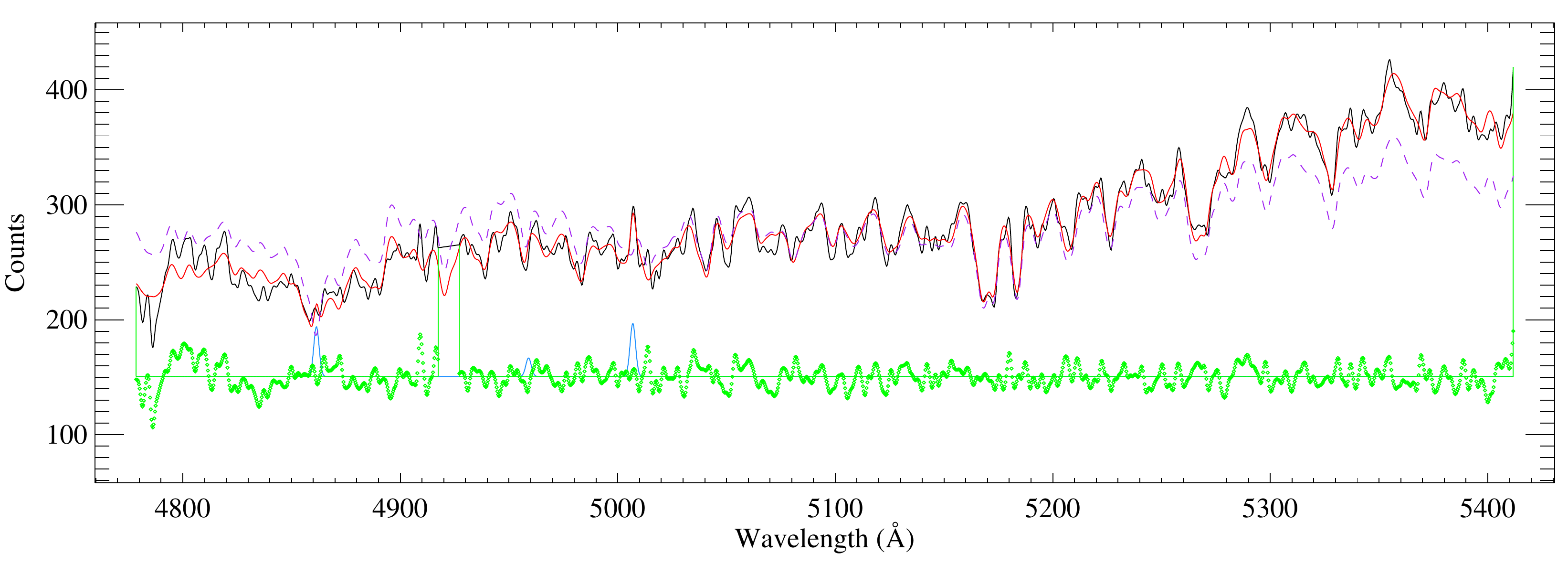}
\caption{Example GANDALF fit for one of the two disc spectra of
NGC~3390. We show the observed spectrum (degraded to the MILES
resolution) in black and the GANDALF best fit in red. The green points
are the difference between the best fit and the observed spectrum,
shifted by an arbitrary positive offset so they appear on the scale. The
best fit is comprised of the H$\beta$ and \oiii\ emission lines (shown
in blue) and a combination of MILES template stars that are convolved by
a LOSVD (purple dashed line) and multiplied by a 6th order polynomial to
account for the relative continuum differences between our setup and the
MILES library. The anomalous feature at $\approx$4920\,\AA\ is due to a
problem with the CCD, and is masked for all analyses.}
\label{fig:gandalf}
\end{figure*}

Cylindrical rotation is therefore not a ubiquitous feature of boxy and
peanut-shaped bulges. There is also some evidence that it is less strong
or even absent in peanut-shaped bulges compared to boxy bulges. It is
worth considering whether this might have a dynamical origin. In
simulations, bars in edge-on galaxies appear peanut-shaped when viewed
with the bar perpendicular to the line-of-sight, and boxy when the bar
is at an intermediate angle \citep[e.g.][]{Combes:1990}. End-on bars
appear neither boxy nor peanut-shaped. The velocity fields of these
simulated galaxies vary with the orientation of the bar in the sense
that the closer the bar is to end-on (i.e. the less strongly
peanut-shaped it appears), the more closely packed are the lines of
constant velocity (e.g. figure 6 of \citealt{Combes:1990} and figure 12
of \citealt{Athanassoula:2002}). There is, however, no clear sign of a
systemic strengthening or weakening of cylindrical rotation as the
viewing angle varies in these published studies. Rather, the strength
and extent of cylindrical rotation in the simulations of
\cite{Athanassoula:2002} is a function of dark halo concentration in the
sense that less concentrated dark haloes host boxy bulges in which
cylindrical rotation is absent or less extended.

On a different note, in ESO~311--G012, NGC~1381, NGC~3390 and NGC~5746
(i.e. all the galaxies with boxy or peanut-shaped bulges except
IC~4767), there is a general downwards trend in velocity dispersion with
increasing height above the disc. This is most clear in ESO~311--G012
and NGC~5746. This behaviour is also seen in classical bulges
(\reffig{fig:sauron}), but the details of the variation of the velocity
dispersion with height were crucial constraints in the Milky Way models
of \cite{Howard:2009} and \cite{Shen:2010}. These authors placed an
upper limit on the possible contribution of any dynamically hot spheroid
component by comparing photometric and kinematic observations to
dynamical models. Their analysis implies that the Milky Way has, at
most, an extremely low-mass classical bulge, i.e. it is a pure disc
galaxy \citep[but see][]{Babusiaux:2010}. Without a comparison to
dynamical models custom-made for each galaxy, which is beyond the scope
of the present work, it is not possible here to draw such a strong
conclusion on the basis of our velocity and dispersion profiles. 

\section{Line strengths and stellar populations}
\label{sec:populations}

\subsection{Methods}
\label{sec:populationmethods}

The spatial variations of stellar population properties offer additional
constraints on the nature and evolution of bulges. We therefore measured
the strengths of the absorption lines present in our data (H$\beta$,
Fe5015, Mg\,$b$, Fe5270, Fe5335 and Fe5406) in the Lick/IDS system
\citep{Burstein:1984,Worthey:1994,Trager:1998} and compared these to
single stellar population (SSP) models, yielding SSP-equivalent
luminosity-weighted ages, metallicities \ZH\ and $\alpha$-element
enhancement \aFe. 

To make these measurements, we require high $S/N$ to reliably subtract
nebular emission and thus accurately measure the absorption line
properties. We therefore binned the major-axis slit into three large bins:
a single bin for the bulge and two disc bins (one from each side of the
galaxy). We collapsed each of the three offset slits into a single bin
dominated by bulge light (rejecting the very small amount of thick disc
light at larger galactocentric radii). This yields one bulge-dominated
measurement from the centre of the bulge, two disc-dominated
measurements along the major axis and three minor-axis bulge
measurements at increasing galactic height. With data at four positions
along the $z$-axis, it is not really possible to quantify vertical gradients
(although as we discuss in the introduction, these are of interest).
However, we can test our naive picture that a boxy bulge should have
\aFe\ similar to that of its host disc.

Of the absorption lines in our data, only H$\beta$ provides a strong
constraint on stellar age that is weakly dependent on metallicity
\citep{Worthey:1994}, but this absorption feature is unfortunately often
filled in by emission. To compare the population properties of galaxy
discs and bulges, we therefore restrict our analysis to those galaxies
whose major-axis disc spectra are either free of emission or of
sufficient $S/N$ to allow reliable subtraction of the H$\beta$
emission using GANDALF \citep{Sarzi:2006}. This is only true for
NGC~1381, which has no emission, and NGC~3390. The analysis that follows
is therefore restricted to those two galaxies only.

We show an example GANDALF fit in \reffig{fig:gandalf}. The best fitting
GANDALF solution is comprised of the H$\beta$ and \oiii\ emission lines
and a combination of flux-calibrated MILES template stars that are
convolved by a LOSVD and multiplied by a 6th order polynomial to account
for the relative continuum differences between our setup and the MILES
library. To produce an emission-cleaned spectrum that can be used to
measure absorption line properties and that can be compared to models,
we subtract the GANDALF emission lines from the observed spectrum and
divide it by the 6th order polynomial, achieving a rough relative flux
calibration. In practice, the polynomial is extremely slowly varying and
close to first order over the bandpasses of the Lick indices, so this
affects the measured line strengths at a level significantly below the
uncertainties due to the noise.

It is then possible to measure the strengths of the lines in the
Lick/IDS system, which we do using the method of \cite{Kuntschner:2000}.
But before these measurements are made, the spectra must be convolved
with a wavelength-dependent Gaussian to account for the difference
between the spectral resolution of our setup and that of the Lick/IDS
system. The resulting indices are further corrected for (1) the
broadening effects of the galaxies' velocity dispersions (which
generally reduce the strength of absorption features) and (2) the small
systematic offsets caused by, e.g., the remaining differences between
the continuum shapes our spectra and the Lick/IDS spectra. We apply the
offsets from the MILES system given in the appendix of
\cite{Paudel:2010}. This offset of course only affects the absolute
value of the indices, not their spatial gradients.

We use the line strengths to measure stellar population properties using
the models of \cite{Thomas:2003}. These models incorporate non-solar
\aFe. The best-fitting SSP-equivalent model for each spectrum is found
by exploring an interpolated grid of model Lick indices in age, \ZH\ and
\aFe\ \citep[see, e.g.,][]{Proctor:2000,Kuntschner:2010}. We exclude
Fe5015 from this analysis. It is still sometimes contaminated by \oiii\
emission, it is close to the bad pixels on the CCD, and it covers an
extremely wide bandpass so is subject to relatively large continuum
systematics and Lick offsets. This line provides very little additional
constraints on the optimal SSP over those offered by the other three Fe
lines.

\subsection{Results}
\label{sec:populationresults}

\begin{figure}
\includegraphics[width=4.15cm]{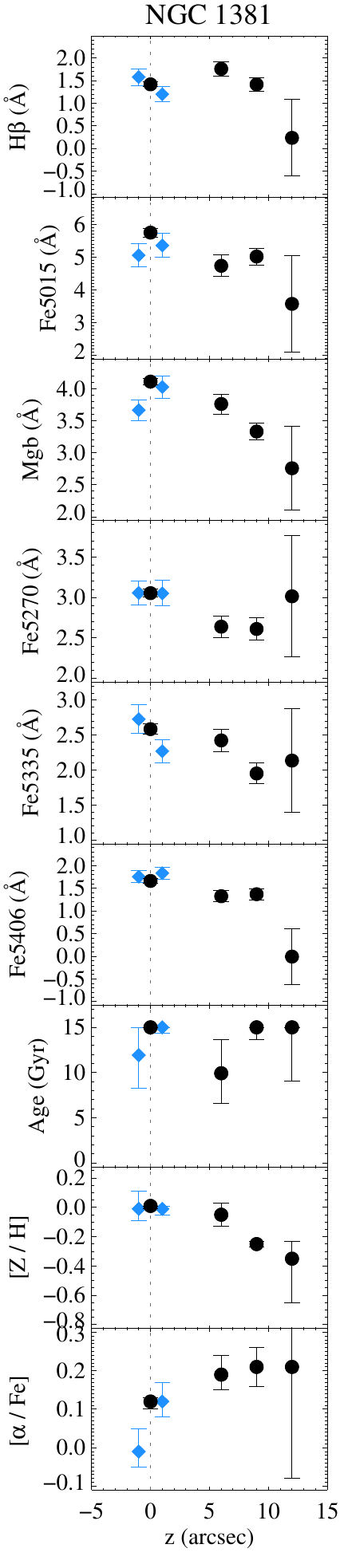}
\includegraphics[width=4.15cm]{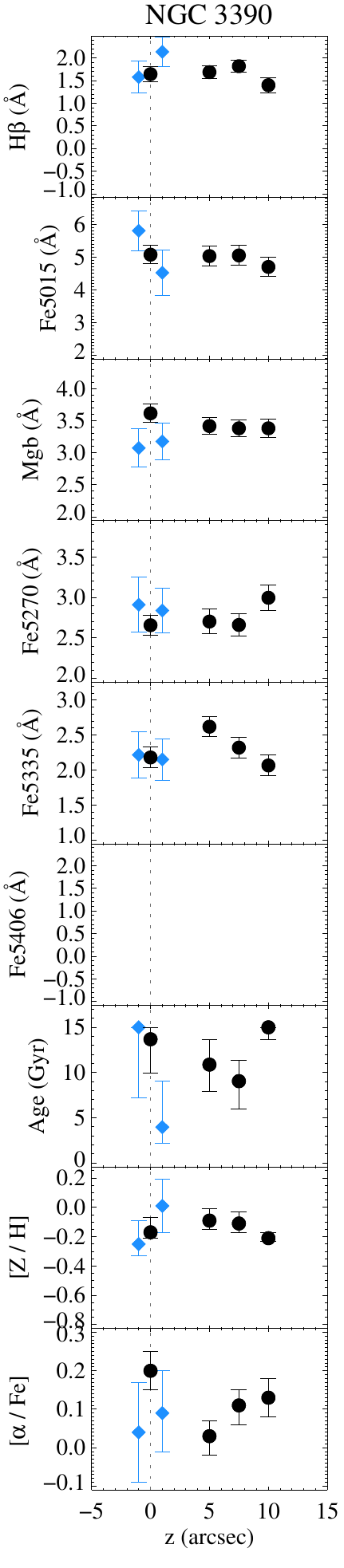}
\caption{Lick indices and luminosity-weighted SSP-equivalent ages,
metallicities and \aFe\ ratios as a function of the perpendicular
distance to the major axis for NGC~1381 and NGC~3390. Measurements are
made at the center of the galaxy (black square at $z$ = 0), two disc
bins along the major axis (blue diamonds) and three points at
increasing distance along the minor axis (black squares at $z \ne 0$).
The disc bins correspond to spectra at $z = 0$, but we added a spatial
offset of $\pm$1\arcsec\ for clarity in these plots. The $z = 12\arcsec$
bulge spectrum in NGC~1381 has some line strengths that are consistent with
an equivalent width of zero, and the uncertainties in \aFe\ for this bin
extend over the full model grid. The position is shown merely for
completeness. The Fe5406 feature is redshifted beyond the observed
wavelength range for NGC~3390 and could
therefore not be measured.}
\label{fig:lick}
\end{figure}

We show the derived Lick indices and corresponding SSP-equivalent
luminosity-weighted age, \ZH\ and \aFe\ as a function of height above
the galactic plane in \reffig{fig:lick}. There are two additional
points on the major axis ($z = 0$), which are the measurements from the
two disc bins.

In both galaxies, there is no sign of a systematic trend in the
age-sensitive H$\beta$ line as a function of $z$, or in the derived age.

In NGC~1381 there are broadly decreasing trends with increasing height
in the metallicity-sensitive Fe and Mg lines. These are qualitatively
similar to the minor-axis trends seen in the SAURON classical bulges
shown in \reffig{fig:sauron} (see \citealt{Kuntschner:2006} for the
corresponding line strength maps). The SSP-equivalent stellar population
properties derived from these line strengths suggest that the bulge is
comprised of an old stellar population, $\gtrsim 10$\,Gyr with no
systematic trend in age as a function of $z$, but there is a clear
decreasing trend in \ZH. We also note that \aFe\ in the bulge at large
$z$ is greater than that in the disc bins (and indeed the central bin of
the major axis). The simplest interpretation of these results is that
NGC~1381 is comprised of a disc and a bulge whose stars formed rapidly.
This explains the general trend in \aFe\ as a function of height, as
disc light (with its lower \aFe) contributes less and less to the
integrated spectrum. In summary, the bulge of NGC~1381 is chemically
similar to that of classical bulges and elliptical galaxies.

Given this classical bulge behaviour, how do we explain the boxy
appearance of the bulge? We speculate that this is a composite bulge:
its chemical properties are explained by the presence of a classical
bulge, but its appearance suggests the simultaneous presence of a bar
(which appears boxy in projection). This is consistent with the stellar
rotation, neither cylindrical nor strongly non-cylindrical. The
double-hump of the rotation curve even hints at the presence of a small
discy pseudobulge. This galaxy thus seems to have all three kinds of
bulges.

The situation in NGC~3390 is much simpler. Within the (not
insignificant) uncertainties, its bulge has no minor-axis metallicity
gradient, and the \aFe\ measurements in the bulge bins at large $z$ are
entirely consistent with those of the disc. That is to say, this
galaxy's bulge is made of material similar to that in its disc, and the
process that has scattered it to large $z$ has not established a
systematic metallicity gradient. This is consistent with the kinematics
of this bulge, which is cylindrically rotating to great precision in all
the slits (including the major axis). The simplest interpretation of
NGC~3390 is therefore that it is a pure disc galaxy.

\section{Discussion}
\label{sec:discussion}

We analysed the kinematics and stellar populations of the boxy or
peanut-shaped bulges of a sample of five edge-on disc galaxies. One boxy
bulge (NGC~3990) rotates perfectly cylindrically, while another
(ESO~311--G012) is very close to cylindrical. It is not possible to
cleanly classify the rotation in the boxy bulge of NGC~1381 or the
peanut-shaped bulge of NGC~5746. The strongly peanut-shaped bulge of
IC~4767 is in normal rotation, the velocity falling systematically with
height. Cylindrical rotation is therefore not a ubiquitous behaviour in
bulges that appear boxy or peanut-shaped. 

Our sample is small and this work is exploratory, but we also tentatively
note that cylindrical rotation may be less strong or even absent in
peanut-shaped bulges, compared to boxy bulges. While this is not a
prediction of simulations, models are able to qualitatively reproduce
a case like IC~4767 by reducing the central concentration of the dark
halo \citep{Athanassoula:2002}.

Low $S/N$ and the presence of emission are less of a problem when
determining absorption line stellar kinematics than when measuring
absorption line strengths. We were therefore only able to measure
minor-axis trends in Lick indices and derive luminosity-weighted
SSP-equivalent age, \ZH\ and \aFe\ for two of the five galaxies. Of
these, NGC~3390 exhibits behaviour consistent with being a pure disc:
there is no significant difference in \ZH\ and \aFe\ between the disc
and the boxy bulge at large galactic heights, suggesting that the boxy
bulge formed from redistributed disc material. NGC~1381, however, shows
a clear negative metallicity gradient as a function of $z$, while \aFe\
in the bulge at large $z$ is significantly enhanced compared to that in
the disc. This behaviour is consistent with a simple disc $+$ classical
bulge model \citep[e.g.][ for the edge-on S0 NGC~3115]{Norris:2006}.
Combined with the marginal cylindrical rotation and the boxy shape of
the bulge, this suggests that the centre of NGC~1381 hosts a composite
bulge comprised of a classical bulge, a discy pseudobulge and a thick
bar.

The naive expectation that we described in our introduction, that a boxy
bulge should have the same \aFe\ as its host disk if it formed through
the bar buckling scenario and is part of a pure disk galaxy, is one part
of our argument that NGC~3390 may be a pure disk galaxy. This picture is
complicated by at least two issues. (1) The disk could have formed
rapidly and then buckled promptly, forming an $\alpha$-enhanced thick
bar, i.e.~a boxy bulge, and (2) In a system made up of stars in which
\aFe\ is anticorrelated with \ZH\ (e.g. as seen in young stars in the
Galaxy, \citealt{Edvardsson:1993}), a negative spatial metallicity
gradient would necessarily coincide with a positive spatial \aFe\
gradient.

Regarding point (1), we note that the `wild disc' scenario (in which
early star formation in the progenitors of disk galaxies takes place on
short timescales in massive clumps, e.g.
\citealt{Noguchi:1999,Elmegreen:2004,Bournaud:2007}) probably cannot
explain how a bar could be $\alpha$-enhanced with respect to its host
disc. In dynamical simulations, these clumps rapidly migrate inward
through dynamical friction and combine to form an axisymmetric feature
that would probably look like an $\alpha$-enhanced classical bulge (both
in terms of populations and dynamics). Chemodynamical simulations of bar
formation are clearly necessary to clarify the issue.

Such studies should also address the more general questions raised by
this work. (1) Is the double hump we see in the rotation curve along the
major axis of two of the boxy and peanut-shaped bulges definitely
associated with a flattened discy pseudobulge? How far does its
influence extend above the plane? (2) How does the viewing angle of the
bar affect the observed kinematics? For example, do we expect weaker
cylindrical rotation in peanut-shaped bulges? This question was
addressed by \cite{Athanassoula:2002}, but quantitative (rather than
qualitative) comparisons to observations are necessary. (3) Do the
minor-axis metallicity and age gradients of simulated bars differ
systematically from those of classical bulges? (4) Can simulated bars
form in a way that explains the presence of a positive minor-axis \aFe\
gradient, as seen in NGC~1381, without the additional presence of a
classical bulge?

The $\alpha$-enhanced bulge of NGC~1381 suggests that not all boxy
bulges are found in pure disc galaxies (although some, including the
Milky Way, may be). Since the prevalence of giant pure disc galaxies is
an acute problem for simulations of galaxy formation in a cosmological
context, it is clear that a combined analysis of the stellar dynamics
and populations of the central regions of galaxies is an extremely
powerful tool.

The sample we have studied in the present work is unusual because it is
dominated by boxy and peanut-shaped bulges. Such galaxies are unique
laboratories to investigate the nature of the bars that drive secular
evolution. Moreover, they raise the possibility that a demographically
significant fraction of giant disk galaxies, making up about half the
population, is composed of pure disc galaxies. We have shown that a
property as simple as the degree of cylindrical rotation varies
significantly among boxy/peanut-shaped bulges, as do the properties of
their stellar populations. The driver of this variation is unknown, but
may be a property as fundamental as dark halo concentration. Clearly, if
we want to further our understanding of the formation of boxy and
peanut-shaped bulges, we need data in larger quantities (to look for
correlations with other galaxy properties such as environment and mass
distribution) and at higher $S/N$ (to more accurately measure stellar
population properties). Two-dimensional data from integral-field units
would also be beneficial, allowing the dynamics to be modelled in
detail.

\section*{Acknowledgements}

We thank Oscar Gonzalez and John Kormendy for useful discussions. MJW is
supported by a European Southern Observatory Studentship and MB by the
STFC rolling grant `Astrophysics at Oxford' (PP/E001114/1).

\label{lastpage}

\end{document}